\documentclass[10pt]{article}
\usepackage{amsmath}
\usepackage{latexsym}
\usepackage{amssymb}
\usepackage{amsfonts}
\usepackage{amsthm}
\usepackage{tikz}
\newcommand*\circled[1]{\tikz[baseline=(char.base)]{\node[shape=circle,draw,inner sep=2pt](char){#1};}}
\title{THE  MATHEMATICAL  FOUNDATIONS  OF  \\  GENERAL  RELATIVITY  REVISITED }
\author{J.F. Pommaret \\ CERMICS, Ecole Nationale des Ponts et
  Chauss\'ees,\\6/8 Av. Blaise Pascal, 77455 Marne-la-Vall\'ee Cedex 02,
  France \\
E-mail: jean-francois.pommaret@wanadoo.fr, pommaret@cermics.enpc.fr \\
URL: http://cermics.enpc.fr/${\sim}$pommaret/home.html }
\date{  }
\begin{document}
\maketitle

\noindent
{\bf  ABSTRACT} :  \\

\noindent
\hspace*{4mm} In 1880 S. Lie (1842-1899) studied the groups of transformations depending on a finite number of parameters and now called {\it Lie groups of transformations}. Ten years later he discovered that these groups are only examples of groups of transformations solutions of linear or nonlinear systems of ordinary differential (OD) or partial differential (PD) equations which may even be of high order and are now called {\it Lie pseudogroups of transformations}. During the next fifty years the latter groups have only been studied by two frenchmen, namely Elie Cartan (1869-1951) who is quite famous today, and Ernest Vessiot (1865-1952) who is almost ignored today. We have proved in many books and papers that {\it the Cartan structure equations have nothing to do with the Vessiot structure equations} still not known today. Accordingly, we prove in the first part of the paper:  \\

\noindent
{\bf FIRST FUNDAMENTAL RESULT}: The quadratic terms appearing in the {\it Riemann tensor} must not be identified with the quadratic terms appearing in the well known {\it Maurer-Cartan equations} for Lie groups and a similar comment can be done for the {\it Weyl tensor}. In particular, {\it curvature+torsion} (Cartan) must not be considered as a generalization of {\it curvature alone} (Vessiot). \\

Though we consider that the first formal work on systems of PD equations is dating back to Maurice Janet (1888-1983) who introduced as early as in 1920 
a {\it differential sequence} now called {\it Janet sequence}, it is only around 1970 that Donald Spencer (1912-2001) developped, in a quite independent way, the formal theory of systems of PD equations in order to study Lie pseudogroups, exactly like E. Cartan did with exterior systems. Nevertheless, all the physicists who tried to understand the only book "Lie Equations "  that he published in 1972 with A. Kumpera, have been stopped by the fact that the examples of the Introduction (Janet sequence) have nothing to do with the core of the book (Spencer sequence). We obtain in the second part of the paper:  \\

\noindent
{\bf SECOND FUNDAMENTAL RESULT}: The {\it Ricci tensor} only depends on the nonlinear transformations (called {\it elations} by Cartan in 1922) that describe the " {\it difference} "  existing between the Weyl group (10 parameters of the Poincar\' e subgroup + 1 dilatation) and the conformal group of 
space-time (15 parameters). It can be defined by a canonical splitting, that is to say without using the indices leading to the standard contraction or trace of the Riemann tensor. Meanwhile, we shall obtain the number of components of the Riemann and Weyl tensors without any combinatoric argument on the exchange of indices. Accordingly, the Spencer sequence for the conformal Killing system and its formal adjoint fully describe the Cosserat/Maxwell/Weyl theory but General Relativity (GR) is not coherent at all with this result. \\

At the same time, mixing commutative algebra (module theory) and homological algebra (extension modules) but always supposing that the reader knows a lot about the work of Spencer, V.P. Palamodov (constant coefficients) and M. Kashiwara (variable coefficients)  developped " {\it algebraic analysis} "  in order to study the formal properties of finitely generated {\it differential modules} that do not depend on their {\it presentation} or even on a corresponding {\it differential resolution}, namely the algebraic analogue of a differential sequence. Finally, we get in the third part of the paper:  \\

\noindent
{\bf THIRD FUNDAMENTAL RESULT}: Contrary to other equations of physics (Cauchy equations, Cosserat equations, Maxwell equations), the Einstein equations cannot be " {\it parametrized} ", that is the generic solution cannot be expressed by means of the derivatives of a certain number of arbitrary potential-like functions, solving therefore negatively a 1000 \$ challenge  proposed by J. Wheeler in 1970. \\

As no one of these results can be obtained without the previous difficult purely mathematical arguments and are thus {\it unavoidable}, the purpose of this paper is to present them for the first time in a rather self-contained and elementary way through explicit basic examples.  \\


\noindent
{\bf INTRODUCTION}:  \\

The purpose of this paper is to present an elementary summary of a few recent results obtained through the application of the formal theory of systems of ordinary differential (OD) or partial differential (PD) equations and Lie pseudogroups in order to revisit the mathematical foundations of General relativity (GR). More elementary engineering examples (elasticity theory, electromagnetism (EM)) will also be considered in order to illustrate the quoted three fundamental results that we shall provide. The paper, based on the material of two lectures given at the department of mathematics of the university of Montpellier 2, France, in may 2013, is divided into three parts corresponding to the different formal methods used.  \\

1) FIRST PART: Lie groups of transformations may be considered as Lie pseudogroups of transformations, that is to say groups of transformations solutions of systems of OD or PD equations, but no action type method can be used as parameters never appear any longer.   \\

2) SECOND PART: The work of Cartan is superseded by the use of the canonical Spencer sequence while the work of Vessiot is superseded by the use of the canonical Janet sequence but the link between these two sequences and thus these two works is not known today.   \\

3) THIRD PART: Using duality theory, the formal adjoint of the Spencer operator for the conformal group of transformations of space-time provides the Cosserat equations, the Maxwell equations and the Weyl equations on equal footing but such a result, even if it allows to unify the finite elements of 
engineering sciences, also leads to contradictions in GR that we shall point out.   \\

The new methods involve tools from differential geometry (jet theory, Spencer operator, $\delta$-cohomology) and homological algebra (diagram chasing, snake theorem, extension modules, double duality). The reader may just have a look to the book ([18], review in Zbl 1079.93001) in order to understand the amount of mathematics needed from many domains. \\

The following diagram summarizes at the same time the historical background and the difficulties presented in the abstract:  \\

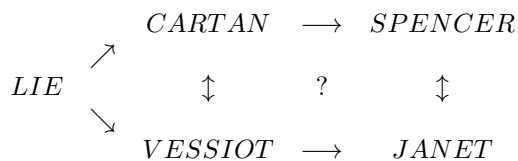
\[   \begin{array}{rcccc}
   &   &  CARTAN   &   \longrightarrow   &  SPENCER  \\
      &   \nearrow  &   &   &   \\
 LIE   &   &   \updownarrow   &  ?  &   \updownarrow   \\
      &   \searrow  &  &  &   \\
   &   &   VESSIOT   &   \longrightarrow   &  JANET   
   \end{array}     \]
\vspace*{1mm}  \\

\noindent
Roughly, Cartan and followers have not been able to " {\it quotient down to the base manifold} " ([1,2]), a result only obtained by Spencer in 1970 through the {\it nonlinear Spencer sequence} ([5],[9],[15],[22]) but in a way quite different from the one followed by Vessiot in 1903 for the same purpose ([17],[25]). Accordingly, the mathematical foundations of mathematical physics must be revisited within this formal framework, though striking it may look like for certain apparently well established theories such as EM and GR.  \\

\noindent
{\bf FIRST PART :  FROM LIE GROUPS TO LIE PSEUDOGROUPS}  \\

If $X$ is a manifold with local coordinates $(x^i)$ for $i=1, ... ,n=dim(X)$, let $\cal{E}$ be a {\it fibered manifold} over $X$, that is a manifold with local coordinates $(x^i,y^k)$ for $i=1,...,n$ and $k=1,...,m$ simply denoted by $(x,y)$, {\it projection} $\pi:{\cal{E}}\rightarrow X:(x,y)\rightarrow (x)$ and changes of local coordinates $\bar{x}=\varphi(x), \bar{y}=\psi(x,y)$. If $\cal{E}$ and $\cal{F}$ are two fibered manifolds over $X$ with respective local coordinates $(x,y)$ and $(x,z)$, we denote by ${\cal{E}}{\times}_X{\cal{F}}$ the {\it fibered product} of $\cal{E}$ and $\cal{F}$ over $X$ as the new fibered manifold over $X$ with local coordinates $(x,y,z)$. We denote by $f:X\rightarrow {\cal{E}}: (x)\rightarrow (x,y=f(x))$ a global {\it section} of $\cal{E}$, that is a map such that $\pi\circ f=id_X$ but local sections over an open set $U\subset X$ may also be considered when needed. We shall use for simplicity the same notation for a fibered manifold and its set of sections while setting $dim_X({\cal{E}})=m$. Under a change of coordinates, a section transforms like $\bar{f}(\varphi(x))=\psi(x,f(x))$ and the derivatives transform like:\\
\[   \frac{\partial{\bar{f}}^l}{\partial{\bar{x}}^r}(\varphi(x)){\partial}_i{\varphi}^r(x)=\frac{\partial{\psi}^l}{\partial x^i}(x,f(x))+\frac{\partial {\psi}^l}{\partial y^k}(x,f(x)){\partial}_if^k(x)  \]
We may introduce new coordinates $(x^i,y^k,y^k_i)$ transforming like:\\
\[ {\bar{y}}^l_r{\partial}_i{\varphi}^r(x)=\frac{\partial{\psi}^l}{\partial x^i}(x,y)+\frac{\partial {\psi}^l}{\partial y^k}(x,y)y^k_i  \]
We shall denote by $J_q({\cal{E}})$ the {\it q-jet bundle} of $\cal{E}$ with local coordinates $(x^i, y^k, y^k_i, y^k_{ij},...)=(x,y_q)$ called {\it jet coordinates} and sections $f_q:(x)\rightarrow (x,f^k(x), f^k_i(x), f^k_{ij}(x), ...)=(x,f_q(x))$ transforming like the sections $j_q(f):(x) \rightarrow (x,f^k(x), {\partial}_if^k(x), {\partial}_{ij}f^k(x), ...)=(x,j_q(f)(x))$ where both $f_q$ and $j_q(f)$ are over the section $f$ of $\cal{E}$. Of course $J_q({\cal{E}})$ is a fibered manifold over $X$ with projection ${\pi}_q$ while $J_{q+r}({\cal{E}})$ is a fibered manifold over $J_q({\cal{E}})$ with projection ${\pi}^{q+r}_q, \forall r\geq 0$.\\

\noindent
{\bf DEFINITION 1.1}: A ({\it nonlinear}) {\it system} of order $q$ on $\cal{E}$ is a fibered submanifold ${\cal{R}}_q\subset J_q({\cal{E}})$ and a {\it solution} of ${\cal{R}}_q$ is a section $f$ of $\cal{E}$ such that $j_q(f)$ is a section of ${\cal{R}}_q$.\\

\noindent
{\bf DEFINITION 1.2}: When the changes of coordinates have the linear form $\bar{x}=\varphi(x),\bar{y}= A(x)y$, we say that $\cal{E}$ is a {\it vector bundle} over $X$. Vector bundles will be denoted by capital letters $C,E,F$ and will have sections denoted by $\xi,\eta,\zeta$. In particular, we shall denote as usual by $T=T(X)$ the {\it tangent bundle} of $X$, by $T^*=T^*(X)$ the {\it cotangent bundle}, by ${\wedge}^rT^*$ the {\it bundle of r-forms} and by $S_qT^*$ the {\it bundle of q-symmetric tensors}. When the changes of coordinates have the form $\bar{x}=\varphi(x),\bar{y}=A(x)y+B(x)$ we say that $\cal{E}$ is an {\it affine bundle} over $X$ and we define the {\it associated vector bundle} $E$ over $X$ by the local coordinates $(x,v)$ changing like $\bar{x}=\varphi(x),\bar{v}=A(x)v$. Finally, If ${\cal{E}}=X\times X$, we shall denote by ${\Pi}_q={\Pi}_q(X,X)$ the open subfibered manifold of $J_q(X\times X)$ defined independently of the coordinate system by $det(y^k_i)\neq 0$ with {\it source projection} ${\alpha}_q:{\Pi}_q\rightarrow X:(x,y_q)\rightarrow (x)$ and {\it target projection} ${\beta}_q:{\Pi}_q\rightarrow X:(x,y_q)\rightarrow (y)$.     \\

\noindent
{\bf DEFINITION 1.3}: If the tangent bundle $T({\cal{E}})$ has local coordinates $(x,y,u,v)$ changing like ${\bar{u}}^j={\partial}_i{\varphi}^j(x)u^i, {\bar{v}}^l=\frac{\partial {\psi}^l}{\partial x^i}(x,y)u^i+\frac{\partial {\psi}^l}{\partial y^k}(x,y)v^k$, we may introduce the {\it vertical bundle} $V({\cal{E}})\subset T({\cal{E}})$ as a vector bundle over $\cal{E}$ with local coordinates $(x,y,v)$ obtained by setting $u=0$ and changes ${\bar{v}}^l=\frac{\partial {\psi}^l}{\partial y^k}(x,y)v^k$. Of course, when $\cal{E}$ is an affine bundle over $X$ with associated vector bundle $E$ over $X$, we have $V({\cal{E}})={\cal{E}}\times_XE$.\\

For a later use, if $\cal{E}$ is a fibered manifold over $X$ and $f$ is a section of $\cal{E}$, we denote by $f^{-1}(V({\cal{E}}))$ the {\it reciprocal image} of $V({\cal{E}})$ by $f$ as the vector bundle over $X$ obtained when replacing $(x,y,v)$ by $(x,f(x),v) $ in each chart. A similar construction may also be done for any affine bundle over ${\cal{E}}$.  \\

We now recall a few basic geometric concepts that will be constantly used through this paper. First of all, if $\xi,\eta\in T$, we define their {\it bracket} $[\xi,\eta]\in T$ by the local formula $([\xi,\eta])^i(x)={\xi}^r(x){\partial}_r{\eta}^i(x)-{\eta}^s(x){\partial}_s{\xi}^i(x)$ leading to the {\it Jacobi identity} $[\xi,[\eta,\zeta]]+[\eta,[\zeta,\xi]]+[\zeta,[\xi,\eta]]=0, \forall \xi,\eta,\zeta \in T$ allowing to define a {\it Lie algebra} and to the useful formula $[T(f)(\xi),T(f)(\eta)]=T(f)([\xi,\eta])$ where $T(f):T(X)\rightarrow T(Y)$ is the tangent mapping of a map $f:X\rightarrow Y$.\\

When $I=\{ i_1< ... < i_r\}$ is a multi-index, we may set $dx^I=dx^{i_1}\wedge ... \wedge dx^{i_r}$ for describing ${\wedge}^rT^*$ by means of a {\it basis} and introduce the {\it exterior derivative} $d:{\wedge}^rT^*\rightarrow {\wedge}^{r+1}T^*:\omega={\omega}_Idx^I \rightarrow d\omega={\partial}_i{\omega}_Idx^i\wedge dx^I$ with $d^2=d\circ d\equiv 0$ in the {\it Poincar\'{e} sequence}:\\
\[  {\wedge}^0T^* \stackrel{d}{\longrightarrow} {\wedge}^1T^* \stackrel{d}{\longrightarrow} {\wedge}^2T^* \stackrel{d}{\longrightarrow} ... \stackrel{d}{\longrightarrow} {\wedge}^nT^* \longrightarrow 0  \]

The {\it Lie derivative} of an $r$-form with respect to a vector field $\xi\in T$ is the linear first order operator ${\cal{L}}(\xi)$ linearly depending on $j_1(\xi)$ and uniquely defined by the following three properties:\\
1) ${\cal{L}}(\xi)f=\xi.f={\xi}^i{\partial}_if, \forall f\in {\wedge}^0T^*=C^{\infty}(X)$.\\
2) ${\cal{L}}(\xi)d=d{\cal{L}}(\xi)$.\\
3) ${\cal{L}}(\xi)(\alpha\wedge \beta)=({\cal{L}}(\xi)\alpha)\wedge \beta+\alpha\wedge ({\cal{L}}(\xi) \beta), \forall \alpha,\beta \in \wedge T^*$.\\
It can be proved that ${\cal{L}}(\xi)=i(\xi)d+di(\xi)$ where $i(\xi)$ is the {\it interior multiplication} $(i(\xi)\omega)_{i_1...i_r}={\xi}^i{\omega}_{ii_1...i_r}$ and that $[{\cal{L}}(\xi),{\cal{L}}(\eta)]={\cal{L}}(\xi)\circ {\cal{L}}(\eta)-{\cal{L}}(\eta)\circ {\cal{L}}(\xi)={\cal{L}}([\xi,\eta]), \forall \xi,\eta\in T$.\\

We now turn to group theory and start with two basic definitions:\\

Let $G$ be a {\it Lie group}, that is a manifold with local coordinates $(a^{\tau})$ for $\tau=1, ... ,p=dim(G)$ called {\it parameters}, a {\it composition} $G\times G \rightarrow G: (a,b)\rightarrow ab$, an {\it inverse} $G \rightarrow G: a \rightarrow a^{-1}$ and an {\it identity} $e\in G$ satisfying:\\
\[(ab)c=a(bc)=abc,\hspace{1cm} aa^{-1}=a^{-1}a=e,\hspace{1cm} ae=ea=a,\hspace{1cm} \forall a,b,c \in G \]

\noindent
{\bf DEFINITION 1.4}: $G$ is said to {\it act} on $X$ if there is a map $X\times G \rightarrow X: (x,a) \rightarrow y=ax=f(x,a)$ such that $(ab)x=a(bx)=abx, \forall a,b\in G, \forall x\in X$ and we shall say that we have a {\it Lie group of transformations} of $X$. In order to simplify the notations, we shall use global notations even if only local actions are existing. It is well known that the action of $G$ onto itself allows to introduce a purely algebraic bracket on its {\it Lie algebra} ${\cal{G}}=T_e(G)$. \\ 

\noindent
{\bf DEFINITION 1.5}: A {\it Lie pseudogroup of transformations} $\Gamma\subset aut(X)$ is a group of transformations solutions of a system of OD or PD equations such that, if $y=f(x)$ and $z=g(y)$ are two solutions, called {\it finite transformations}, that can be composed, then $z=g\circ f(x)=h(x)$ and $x=f^{-1}(y)=g(y)$ are also solutions while $y=x$ is the {\it identity} solution denoted by $id=id_X$ and we shall set $id_q=j_q(id)$.  In all the sequel we shall suppose that $\Gamma$ is {\it transitive} that is $\forall x,y\in X, \exists f\in \Gamma, y=f(x)$ \\

It becomes clear that Lie groups of transformations are particular cases of Lie pseudogroups of transformations as the system defining the finite transformations can be obtained by eliminating the parameters among the equations $y_q=j_q(f)(x,a)$ when $q$ is large enough. The underlying system may be nonlinear and of high order. Looking for transformations "close" to the identity, that is setting $y=x+t\xi(x)+...$ when $t\ll 1$ is a small constant parameter and passing to the limit $t\rightarrow 0$, we may linearize the above nonlinear {\it system of finite Lie equations} in order to obtain a linear {\it system of infinitesimal Lie equations} of the same order for vector fields. Such a system has the property that, if $\xi,\eta$ are two solutions, then $[\xi,\eta]$ is also a solution. Accordingly, the set $\Theta\subset T$ of solutions of this new system satisfies $[\Theta,\Theta]\subset \Theta$ and can therefore be considered as the Lie algebra of $\Gamma$.\\

\noindent
{\bf EXAMPLE 1.6}:  While the {\it affine} transformations $y=ax+b$ are solutions of the second order linear system $y_{xx}=0$, the {\it projective} transformations $y=(ax+b)/(cx+d)$ are solutions of the third order nonlinear system $\Psi\equiv (y_{xxx}/y_x)-\frac{3}{2}(y_{xx}/y_x)^2=0$. The sections of the corresponding linearized systems are respectively satisfying ${\xi}_{xx}=0$ and ${\xi}_{xxx}=0$. The generating differential invariant $\Phi\equiv y_{xx}/y_x$ of the affine case transforms like $u=\bar{u}{\partial}_xf+({\partial}_{xx}f/{\partial}_xf)$ when $\bar{x}=f(x)$ and we let the reader exhibit the corresponding change for $\Psi$ as an exercise. \\

We now sketch the discovery of Vessiot ([17],[25]) {\it still not known today after more than a century} for reasons which are not scientific at all. 
Roughly, a Lie pseudogroup $\Gamma \subset aut(X)$ is made by finite transformations $y=f(x)$ solutions of a (possibly nonlinear) system ${\cal{R}}_q\subset {\Pi}_q$ while the infinitesimal transformations $\xi\in \Theta$ are solutions of the linearized system $R_q=id_q^{-1}(V({\cal{R}}_q))\subset J_q(T)$ as we have $T=id^{-1}(V(X\times X)$. When $\Gamma$ is transitive, there is a canonical epimorphism ${\pi}^q_0:R_q\rightarrow T$. Also, as changes of source $x$ commute with changes of target $y$, they exchange between themselves any generating set of {\it differential invariants} $\{{\Phi}^{\tau}(y_q)\}$ as in the previous example.Then one can introduce a {\it natural bundle} $\cal{F}$ over $X$, also called {\it bundle of geomeric objects}, by patching changes of coordinates of the form $\bar{x}=f(x), \bar{u}=\lambda(u,j_q(f(x))$ thus obtained. A section $\omega$ of $\cal{F}$ is called a {\it geometric object} or {\it structure} on $X$ and transforms like ${\bar{\omega}}(f(x))=\lambda(\omega(x),j_q(f)(x))$ or simply $\bar{\omega}=j_q(f)^{-1}(\omega)$. This is a way to generalize vectors and tensors ($q=1$) or even connections ($q=2$). As a byproduct we have $\Gamma=\{f\in aut(X){\mid} j_q(f)^{-1}(\omega)=\omega\}$ and we may say that $\Gamma$ {\it preserves} $\omega$. Replacing $j_q(f)$ by $f_q$, we also obtain ${\cal{R}}_q=\{f_q\in {\Pi}_q{\mid} f_q^{-1}(\omega)=\omega\}$. Coming back to the infinitesimal point of view and setting $f_t=exp(t\xi)\in aut(X), \forall \xi\in T$, we may define the {\it ordinary Lie derivative} with value in the vector bundle $F_0={\omega}^{-1}(V({\cal{F}}))$ by the formula :\\
\[ {\cal{D}}\xi={\cal{L}}(\xi)\omega=\frac{d}{dt}j_q(f_t)^{-1}(\omega){\mid}_{t=0} \Rightarrow \Theta=\{\xi\in T{\mid}{\cal{L}}(\xi)\omega=0\}  \]
and we say that $\cal{D}$ is a {\it Lie operator} because ${\cal{D}}\xi=0,{\cal{D}}\eta=0\Rightarrow {\cal{D}}[\xi,\eta]=0$ as we already saw.\\

Differentiating $r$ times the equations of $R_q$ that only depend on $j_1(\omega)$, we may obtain the $r$-{\it prolongation} $R_{q+r}=J_r(R_q)\cap J_{q+r}(T)\subset J_r(J_q(T))$. The problem is then to know under what conditions on $\omega$ all the equations of order $q+r$ are obtained by $r$ prolongations only, $\forall r\geq 0$ or, equivalently, $R_q$ is {\it formally integrable} (FI). The solution, found by Vessiot, has been to exhibit another natural vector bundle ${\cal{F}}_1$ with local coordinates $(x,u,v)$ over ${\cal{F}}$ with local coordinates $(x,u)$ and to prove that an {\it equivariant section} $c:{\cal{F}}\rightarrow {\cal{F}}_1:(x,u)\rightarrow (x,u,v=c(u))$ only depends on a finite number of constants called {\it structure constants}. The {\it integrability conditions} (IC) of $R_q$, called {\it Vessiot structure equations}, are of the form $I(j_1(\omega))=c(\omega)$ and are invariant under any change of source.  \\

We provide in a self-contained way and parallel manners the following five striking examples which are among the best nontrivial ones we know and invite the reader to imagine at this stage any possible link that could exist between them (A few specific definitions will be given later on).\\

\noindent
{\bf EXAMPLE 1.7}: Coming back to the last example, we show that Vessiot structure equations may even exist when $n=1$. For this, if $\gamma$ is the geometric object of the affine group $y=ax+b$ and $0\neq \alpha=\alpha (x)dx \in T^*$ is a $1$-form, we consider the object $\omega=(\alpha,\gamma)$ and get at once the two {\it second order Medolaghi equations}:\\  
\[  {\cal{L}}(\xi)\alpha\equiv \alpha {\partial}_x\xi + \xi {\partial}_x\alpha =0, \hspace{1cm} {\cal{L}}(\xi)\gamma\equiv {\partial}_{xx}\xi+\gamma {\partial}_x\xi+ \xi {\partial}_x\gamma =0  \]
Differentiating the first equation and substituting the second, we get the zero order equation:  \\
\[  \xi (\alpha {\partial}_{xx}\alpha-2({\partial}_x\alpha)^2+\alpha \gamma {\partial}_x\alpha-{\alpha}^2{\partial}_x\gamma)=0\hspace{5mm} \Leftrightarrow \hspace{5mm} \xi {\partial}_x(\frac{{\partial}_x\alpha}{{\alpha}^2} - \frac{\gamma}{\alpha} )=0  \]
and the {\it Vessiot structure equation} ${\partial}_x\alpha-\gamma \alpha=c{\alpha}^2$. Alternatively, setting $\beta=-1/\alpha\in T$, we get ${\partial}_x\beta+\gamma\beta=c$. With $\alpha=1, \beta=-1, \gamma=0 \Rightarrow c=0$ we get the translation subgroup $y=x+b$ while, with $\alpha=1/x, \beta=-x, \gamma=0 \Rightarrow c=-1$ we get the dilatation subgroup $y=ax$.  \\

\noindent
{\bf EXAMPLE 1.8}: ({\it Principal homogeneous structure}) When $\Gamma$ is the Lie group of transformations made by the constant translations  $y^i=x^i+a^i$ for $i=1,...,n$ of a manifold $X$ with $dim(X)=n$, the characteristic object invariant by $\Gamma$ is a family $\omega=({\omega}^{\tau}={\omega}^{\tau}_idx^i)\in T^*\times_X...\times_XT^*$ of $n$ $1$-forms with $det(\omega)\neq 0$ in such a way that $\Gamma =\{f\in aut(X){\mid}j_1(f)^{-1}(\omega)=\omega\}$ where $aut(X)$ denotes the pseudogroup of local diffeomorphisms of $X$, $j_q(f)$ denotes the derivatives of $f$ up to order $q$ and $j_1(f)$ acts in the usual way on covariant tensors. For any vector field $\xi\in T=T(X)$ the tangent bundle to $X$, introducing the standard Lie derivative ${\cal{L}}(\xi)$ of forms with respect to $\xi$, we may consider the $n^2$ 
{\it  first order Medolaghi equations}:  \\
\[   {\Omega}^{\tau}_i\equiv ({\cal{L}}(\xi)\omega)^{\tau}_i\equiv {\omega}^{\tau}_r(x){\partial}_i{\xi}^r+{\xi}^r{\partial}_r{\omega}^{\tau}_i(x)=0  \]
The particular situation is found with the special choice $\omega=(dx^i)$ that leads to the involutive system ${\partial}_i{\xi}^k=0$. Introducing the inverse matrix $\alpha=({\alpha}^i_{\tau})={\omega}^{-1}$, the above equations amount to the bracket relations $[\xi,{\alpha}_{\tau}]=0$ and, using crossed derivatives on the {\it solved form} ${\partial}_i{\xi}^k+{\xi}^r{\alpha}^k_{\tau}(x){\partial}_r{\omega}^{\tau}_i(x)=0$, we obtain the $n^2(n-1)/2$ zero order equations:  \\
\[   {\xi}^r{\partial}_r({\alpha}^i_{\rho}(x){\alpha}^j_{\sigma}(x)({\partial}_i{\omega}^{\tau}_j(x)-{\partial}_j{\omega}^{\tau}_i(x)))=0   \]
The {\it integrability conditions} (IC), that is the conditions under which these equations do not bring new equations, are thus the $n^2(n-1)/2$ {\it Vessiot structure equations}:  \\
\[    {\partial}_i{\omega}^{\tau}_j(x)-{\partial}_j{\omega}^{\tau}_i(x)=c^{\tau}_{\rho\sigma}{\omega}^{\rho}_i(x){\omega}^{\sigma}_j(x)   \]
with $n^2(n-1)/2$ {\it structure constants} $c=(c^{\tau}_{\rho\sigma}=-c^{\tau}_{\sigma\rho})$. When $X=G$, these equations can be identified with the {\it Maurer-Cartan equations} (MC) existing in the theory of Lie groups, on the condition to change the sign of the structure constants involved because we have $[{\alpha}_{\rho},{\alpha}_{\sigma}]= - c^{\tau}_{\rho\sigma}{\alpha}_{\tau}$. Writing these equations in the form of the exterior system $d{\omega}^{\tau}=c^{\tau}_{\rho\sigma}{\omega}^{\rho}\wedge{\omega}^{\sigma}$ and closing this system by applying once more the exterior derivative $d$, we obtain the quadratic IC:   \\
\[   c^{\lambda}_{\mu\rho}c^{\mu}_{\sigma\tau}+c^{\lambda}_{\mu\sigma}c^{\mu}_{\tau\rho}+c^{\lambda}_{\mu\tau}c^{\mu}_{\rho\sigma}=0  \]
also called {\it Jacobi relations} $J(c)=0$. \\

\noindent
{\bf EXAMPLE 1.9}: ({\it Riemann structure}) If $\omega=({\omega}_{ij}={\omega}_{ji})\in S_2T^*$ is a metric on a manifold $X$ with $dim(X)=n$ such that $det(\omega)\neq 0$, the Lie pseudogroup of transformations preserving $\omega$ is $\Gamma=\{f\in aut(X){\mid}j_1(f)^{-1}(\omega)=\omega \}$ and is a Lie group with a maximum number of $n(n+1)/2$ parameters. A special metric could be the Euclidean metric when $n=1,2,3$ as in elasticity theory or the Minkowski metric when $n=4$ as in special relativity ([12]). The 
{\it  first order Medolaghi equations}:\\
\[ {\Omega}_{ij}\equiv ({\cal{L}}(\xi)\omega)_{ij}\equiv {\omega}_{rj}(x){\partial}_i{\xi}^r+{\omega}_{ir}(x){\partial}_j{\xi}^r+{\xi}^r{\partial}_r{\omega}_{ij}(x)=0 \]
are also called {\it classical Killing equations} for historical reasons. The main problem is that {\it this system is not involutive} unless we prolong it to order two by differentiating once the equations. For such a purpose, introducing ${\omega}^{-1}=({\omega}^{ij})$ as usual, we may define the {\it Christoffel symbols}:\\
\[ {\gamma}^k_{ij}(x)=\frac{1}{2}{\omega}^{kr}(x)({\partial}_i{\omega}_{rj}(x) +{\partial}_j  {\omega}_{ri}(x) -{\partial}_r{\omega}_{ij}(x))=
{\gamma}^k_{ji}(x) \]
This is a new geometric object of order $2$ providing the Levi-Civita isomorphism $j_1(\omega)=(\omega,\partial \omega)\simeq (\omega,\gamma)$ of affine bundles and allowing to obtain the {\it second order Medolaghi equations}:\\
\[ {\Gamma}^k_{ij}\equiv ({\cal{L}}(\xi)\gamma)^k_{ij}\equiv {\partial}_{ij}{\xi}^k+{\gamma}^k_{rj}(x){\partial}_i{\xi}^r+{\gamma}^k_{ir}(x){\partial}_j{\xi}^r-{\gamma}^r_{ij}(x){\partial}_r{\xi}^k+{\xi}^r{\partial}_r{\gamma}^k_{ij}(x)=0   \]
Surprisingly, the following expression called {\it Riemann tensor}:\\
\[ {\rho}^k_{lij}(x)\equiv {\partial}_i{\gamma}^k_{lj}(x)-{\partial}_j{\gamma}^k_{li}(x)+{\gamma}^r_{lj}(x){\gamma}^k_{ri}(x)-{\gamma}^r_{li}(x){\gamma}^k_{rj}(x)  \]
is still a first order geometric object and even a $4$-tensor with $n^2(n^2-1)/12$ independent components satisfying the purely algebraic relations :\\
 \[   {\rho}^k_{lij}+{\rho}^k_{ijl}+{\rho}^k_{jli}=0, \hspace{5mm} {\omega}_{rl}{\rho}^l_{kij}+{\omega}_{kr}{\rho}^r_{lij}=0  \]
Accordingly, the IC must express that the new first order equations $R^k_{lij}\equiv ({\cal{L}}(\xi)\rho)^k_{lij}=0$ are only linear combinations of the previous ones and we get the {\it Vessiot structure equations}:\\
    \[  {\rho}^k_{lij}(x)=c({\delta}^k_i{\omega}_{lj}(x)-{\delta}^k_j{\omega}_{li}(x))   \]
with the only {\it structure constant} $c$ describing the constant Riemannian curvature condition of Eisenhart ([4],[16,p139]). One can proceed similarly for the {\it conformal Killing system} ${\cal{L}}(\xi)\omega=A(x)\omega$ and obtain that the {\it Weyl tensor} must vanish, without any structure constant involved 
([16,p 141]).  \\ 

\noindent
{\bf EXAMPLE 1.10}: ({\it Contact structure}) We only treat the case $dim(X)=3$ as the case $dim(X)={2p+1}$ needs much more work ([15,p684]). Let us consider the so-called {\it contact} $1$-form $\alpha=dx^1-x^3dx^2$ and consider the Lie pseudogroup $\Gamma\subset aut(X)$ of (local) transformations preserving $\alpha$ up to a function factor, that is $\Gamma=\{f\in aut(X){\mid}j_1(f)^{-1}(\alpha)= \rho \alpha\}$ where again $j_q(f)$ is a symbolic way for writing out the derivatives of $f$ up to order $q$ and $\alpha$ transforms like a $1$-covariant tensor. It may be tempting to look for a kind of "{\it object} " the invariance of which should characterize $\Gamma$. Introducing the exterior derivative $d\alpha=dx^2\wedge dx^3$ as a $2$-form, we obtain the volume $3$-form $\alpha\wedge d\alpha=dx^1\wedge dx^2\wedge dx^3$. As it is well known that the exterior derivative commutes with any diffeomorphism, we obtain sucessively:\\
\[      j_1(f)^{-1}(d\alpha)=d(j_1(f)^{-1}(\alpha))=d(\rho \alpha)=\rho d\alpha +d\rho \wedge \alpha  \Rightarrow j_1(f)^{-1}(\alpha \wedge d\alpha)={\rho}^2(\alpha\wedge d\alpha)   \]
As the volume $3$-form $\alpha\wedge d\alpha$ transforms through a division by the Jacobian determinant $\Delta=\partial (f^1,f^2,f^3)/\partial (x^1,x^2,x^3)\neq 0$ of the transformation $y=f(x)$ with inverse $x=f^{-1}(y)=g(y)$, {\it the desired object is thus no longer a} $1$-{\it form but a} $1$-{\it form density} $\omega=({\omega}_1,{\omega}_2,{\omega}_3)$ transforming like a $1$-form but up to a division by the square root of the Jacobian determinant. It follows that the infinitesimal contact transformations are vector fields $\xi\in T=T(X)$ the tangent bundle of $X$, satisfying the $3$ so-called 
{\it  first order Medolaghi equations}:ÊÊÊ\\
\[  {\Omega}_i\equiv ({\cal{L}}(\xi)\omega)_i\equiv {\omega}_r(x){\partial}_i{\xi}^r-(1/2){\omega}_i(x){\partial}_r{\xi}^r+{\xi}^r{\partial}_r{\omega}_i(x)=0   \]
When $\omega=(1,-x^3,0)$, we obtain the {\it special} involutive system:ÊÊ\\
\[  {\partial}_3{\xi}^3+{\partial}_2{\xi}^2+2x^3{\partial}_1{\xi}^2-{\partial}_1{\xi}^1=0, {\partial}_3{\xi}^1-x^3{\partial}_3{\xi}^2=0, {\partial}_2{\xi}^1-x^3{\partial}_2{\xi}^2+x^3{\partial}_1{\xi}^1-(x^3)^2{\partial}_1{\xi}^2-{\xi}^3=0   \]
with  $2$ equations of class $3$ and $1$ equation of class 2 obtained by exchanging $x^1$ and $x^3$ (see later on for a precise definition) and thus only $1$ {\it compatibility conditions} (CC) for the second members.\\
For an arbitrary $\omega$, we may ask about the differential conditions on $\omega$ such that all the equations of order $r+1$ are only obtained by differentiating $r$ times the first order equations, exactly like in the special situation just considered where the system is involutive. We notice that, in a symbolic way, $\omega \wedge d\omega$ is now a scalar $c(x)$ providing the zero order equation ${\xi}^r{\partial}_rc(x)=0$ and the condition is $c(x)=c=cst$. The {\it integrability condition} (IC) is the {\it Vessiot structure equation}:   \\
\[  {\omega}_1({\partial}_2{\omega}_3-{\partial}_3{\omega}_2)+{\omega}_2({\partial}_3{\omega}_1-{\partial}_1{\omega}_3)+
{\omega}_3({\partial}_1{\omega}_2-{\partial}_2{\omega}_1)=c    \]
involving the only {\it structure constant} $c$.\\
For $\omega=(1,-x^3,0)$, we get $c=1$. If we choose $\bar{\omega}=(1,0,0)$ leading to $\bar{c}=0$, we may define $\bar{\Gamma}=\{f\in aut(X){\mid} j_1(f)^{-1}(\bar{\omega})=\bar{\omega}\}$ with infinitesimal transformations satisfying the involutive system:\\
\[  {\partial}_3{\xi}^3+{\partial}_2{\xi}^2-{\partial}_1{\xi}^1=0, {\partial}_3{\xi}^1=0, {\partial}_2{\xi}^1=0  \]
with again $2$ equations of class $3$ and $1$ equation of class $2$. \\

\noindent
{\bf EXAMPLE 1.11}: ({\it Unimodular contact structure})  With similar notations, let us again set $\alpha=dx^1-x^3dx^2\Rightarrow d\alpha=dx^2\wedge dx^3$  but let us now consider the new Lie pseudogroup  of transformations preserving $\alpha$ and thus $d\alpha$ too, that is preserving the {\it mixed} object $\omega=(\alpha,\beta)\in T^*\times_X{\wedge}^2T^*$ made up by a $1$-form $\alpha$ and a $2$-form $\beta$ with 
$\gamma = \alpha\wedge \beta\neq 0$ and $d\alpha=\beta\Rightarrow d\beta=0$. Then $\Gamma$ is a Lie subpseudogroup of the one just considered in the previous example and the corresponding infinitesimal transformations now satisfy the involutive system:\\
\[   {\partial}_1{\xi}^1=0, {\partial}_1{\xi}^2=0, {\partial}_1{\xi}^3=0, {\partial}_2{\xi}^1+x^3{\partial}_3{\xi}^3-{\xi}^3=0, {\partial}_2{\xi}^2+{\partial}_3{\xi}^3=0, {\partial}_3{\xi}^1-x^3{\partial}_3{\xi}^2=0  \]
with $3$ equations of class $3$, $2$ equations of class $2$ and $1$ equation of class $1$ if we exchange $x^1$ with $x^3$, a result leading now to $4$ CC. \\
More generally, when $\omega=(\alpha,\beta)$ where $\alpha$ is a $1$-form and $\beta$ is a $2$-form satifying $\alpha\wedge \beta\neq 0$, {\it we may study the same problem as before} for the {\it general} system ${\cal{L}}(\xi)\alpha=0, {\cal{L}}(\xi)\beta=0$. We let the reader provide the details of the tedious computation involved as it is at this point that computer algebra may be used ([11]). The result, not evident at first sight, is that the $2$-form $d\alpha$ {\it must} be proportional to the $2$-form $\beta$, that is $d\alpha=c'(x)\beta$ and thus $\alpha\wedge d\alpha=c'(x)\alpha\wedge\beta$. As $\alpha\wedge\beta\neq 0$, we {\it must} have $c'(x)=c'=cst$ and thus $d\alpha=c'\beta$. Similarly, we get $d\beta=c''\alpha\wedge\beta$ and obtain finally the $4$ {\it Vessiot structure equations} $d\alpha=c'\beta,d\beta=c''\alpha\wedge\beta$ involving $2$ {\it structure constants} $c=(c',c'')$. Contrary to the previous situation (but like in the Riemann case !) we notice that we have now $2$ structure equations not containing any constant (called {\it first kind} by Vessiot) and $2$ structure equations with the same number of different constants (called {\it second kind} by Vessiot), namely $\alpha\wedge d\alpha=c'\alpha\wedge\beta, d\beta=c''\alpha\wedge\beta$.\\
Finally, closing this system by taking once more the exterior derivative, we get $0=d^2\alpha=c'd\beta=c'c''\alpha\wedge\beta$ and thus the unexpected purely algebraic {\it Jacobi condition} $c'c''=0$. For the special choice $\omega=(dx^1-x^3dx^2, dx^2\wedge dx^3)$ we get $c=(1,0)$, for the second special choice $\bar{\omega}=(dx^1,dx^2\wedge dx^3)$ we get $\bar{c}=(0,0)$ and for the third special choice $\bar{\bar{\omega}}=((1/x^1)dx^1, x^1dx^2\wedge dx^3)$ we get $\bar{\bar{c}}=(0,1)$.  \\

\noindent
{\bf FIRST FUNDAMENTAL RESULT }: Comparing the various Vessiot structure equations containing structure constants that we have just presented and that we recall below in a symbolic way, we notice that {\it these structure constants are absolutely on equal footing} though they have in general nothing to do with any Lie algebra.\\
\[   \hspace{2mm}   \left\{
\begin{array}{rcl}
\partial \alpha -\gamma \alpha & = & c \hspace{2mm}{\alpha}^2  \\
\partial \beta + \gamma \beta & = & c
\end{array}
\right.     \]
 \[\begin{array}{rcl}
\partial \omega - \partial \omega & = & c \hspace{2mm} \omega \hspace{1mm} \omega   \\
\partial \gamma -\partial \gamma + \gamma \gamma -\gamma \gamma & = & c \hspace{2mm}(\delta \omega-\delta \omega)  \\
\omega \wedge (\partial \omega - \partial \omega) & = & c 
\end{array}  \]
\[ \hspace{18mm} \left\{
\begin{array}{rcl}
d \alpha & = & c'\hspace{2mm} \beta  \\
d \beta & = & c'' \hspace{2mm}\alpha \wedge \beta
\end{array}  
\right.    \]
Accordingly, the fact that the ones appearing in the MC equations are related to a Lie algebra is a coincidence and {\it the Cartan structure equations have nothing to do with the Vessiot structure equations}. Also, as their factors are either constant, linear or quadratic, {\it any identification of the quadratic terms appearing in the Riemann tensor with the quadratic terms appearing in the MC equations is definitively not correct} ([22]). We also understand why the torsion is {\it automatically combined} with curvature in the Cartan structure equations but {\it totally absent} from the Vessiot structure equations, even though the underlying group (translations + rotations) is the same.   \\

\noindent
{\bf HISTORICAL REMARK 1.12}: Despite the prophetic comments of the italian mathematician Ugo Amaldi in 1909 ([16,p46-52]), it has been a pity that Cartan deliberately ignored the work of Vessiot at the beginning of the last century and that the things did not improve afterwards in the eighties with Spencer and coworkers (Compare MR 720863 (85m:12004) and MR 954613 (90e:58166)).\\

\noindent
{\bf SECOND PART : THE JANET AND SPENCER SEQUENCES}   \\

Let $\mu=({\mu}_1,...,{\mu}_n)$ be a multi-index with {\it length} ${\mid}\mu{\mid}={\mu}_1+...+{\mu}_n$, {\it class} $i$ if ${\mu}_1=...={\mu}_{i-1}=0,{\mu}_i\neq 0$ and $\mu +1_i=({\mu}_1,...,{\mu}_{i-1},{\mu}_i +1, {\mu}_{i+1},...,{\mu}_n)$. We set $y_q=\{y^k_{\mu}{\mid} 1\leq k\leq m, 0\leq {\mid}\mu{\mid}\leq q\}$ with $y^k_{\mu}=y^k$ when ${\mid}\mu{\mid}=0$. If $E$ is a vector bundle over $X$ with local coordinates $(x,y)$ and $J_q(E)$ is the $q$-{\it jet bundle} of $E$ with local coordinates $(x,y_q)$, the {\it Spencer operator} just allows to distinguish a section ${\xi}_q$ from a section $j_q(\xi)$ by introducing a kind of "{\it difference}" through the operator $D:J_{q+1}(E)\rightarrow T^*\otimes J_q(E): {\xi}_{q+1}\rightarrow j_1({\xi}_q)-{\xi}_{q+1}$ with local components $({\partial}_i{\xi}^k(x)-{\xi}^k_i(x), {\partial}_i{\xi}^k_j(x)-{\xi}^k_{ij}(x),...) $ and more generally $(D{\xi}_{q+1})^k_{\mu,i}(x)={\partial}_i{\xi}^k_{\mu}(x)-{\xi}^k_{\mu+1_i}(x)$. Minus the restriction of $D$ to the kernel $S_{q+1}T^*\otimes E$ of the canonical projection ${\pi}^{q+1}_q:J_{q+1}(E)\rightarrow J_q(E)$ can be extended to the  {\it Spencer map} $\delta:{\wedge}^sT^*\otimes S_{q+1}T^*\otimes E\rightarrow {\wedge}^{s+1}T^*\otimes S_qT^*\otimes E$ defined by $({\delta \omega})^k_{\mu}=dx^i\wedge {\omega}^k_{\mu+1_i}$. The kernel of $D$ is made by sections such that ${\xi}_{q+1}=j_1({\xi}_q)=j_2({\xi}_{q-1})=...=j_{q+1}({\xi})$. Finally, if $R_q\subset J_q(E)$ is a {\it system} of order $q$ on $E$ locally defined by linear equations ${\Phi}^{\tau}(x,y_q)\equiv a^{\tau\mu}_k(x)y^k_{\mu}=0$, the $r$-{\it prolongation} $R_{q+r}={\rho}_r(R_q)=J_r(R_q)\cap J_{q+r}(E)\subset J_r(J_q(E))$ is locally defined when $r=1$ by the linear equations ${\Phi}^{\tau}(x,y_q)=0, d_i{\Phi}^{\tau}(x,y_{q+1})\equiv a^{\tau\mu}_k(x)y^k_{\mu+1_i}+{\partial}_ia^{\tau\mu}_k(x)y^k_{\mu}=0$ and has {\it symbol} $g_{q+r}=R_{q+r}\cap S_{q+r}T^*\otimes E\subset J_{q+r}(E)$ locally defined by $a^{\tau\mu}_k(x){\xi}^k_{\mu+\nu}=0, \mid\mu\mid=q,\mid\nu\mid=r$ if one looks at the {\it top order terms}. If ${\xi}_{q+1}\in R_{q+1}$ is over ${\xi}_q\in R_q$, differentiating the identity $a^{\tau\mu}_k(x){\xi}^k_{\mu}(x)\equiv 0$ with respect to $x^i$ and substracting the identity $a^{\tau\mu}_k(x){\xi}^k_{\mu+1_i}(x)+{\partial}_ia^{\tau\mu}_k(x){\xi}^k_{\mu}(x)\equiv 0$, we obtain the identity $a^{\tau\mu}_k(x)({\partial}_i{\xi}^k_{\mu}(x)-{\xi}^k_{\mu+1_i}(x))\equiv 0$ and thus the restriction $D:R_{q+1}\rightarrow T^*\otimes R_q$. This first order operator induces, up to sign, the purely algebraic monomorphism $ 0 \rightarrow g_{q+1} \stackrel{\delta}{\rightarrow} T^*\otimes g_q$ on the symbol level ([17],[24). {\it The Spencer operator has never been used in} GR.   \\
    
\noindent
{\bf DEFINITION 2.1}: $R_q$ is said to be {\it formally integrable} (FI) when the restriction ${\pi}^{q+r+1}_{q+r}:R_{q+r+1}\rightarrow R_{q+r} $ is an epimorphism $\forall r\geq 0$. In that case, the {\it Spencer form} $R_{q+1}\subset J_1(R_q)$ is a canonical equivalent formally integrable first order system on $R_q$ with no zero order equations.\\

\noindent
{\bf DEFINITION 2.2}: $R_q$ is said to be {\it involutive} when it is formally integrable and the symbol $g_q$ is {\it involutive}, that is all the sequences $... \stackrel{\delta}{\rightarrow} {\wedge}^sT^*\otimes g_{q+r}\stackrel{\delta}{\rightarrow}...$ are exact $\forall 0\leq s\leq n, \forall r\geq 0$. Equivalently, using a linear change of local coordinates if necessary, we may {\it successively} solve the maximum number ${\beta}^n_q, {\beta}^{n-1}_q, ... , {\beta}^1_q$ of equations with respect to the {\it leading} or {\it principal} jet coordinates of strict order $q$ and class $n,n-1,...,1$. Then $R_q$ is involutive if $R_{q+1}$ is obtained by only prolonging the ${\beta}^i_q$ equations of class $i$ with respect to $d_1,...,d_i$ for $i=1,...,n$. In that case, such a prolongation procedure allows to compute {\it in a unique way} the principal jets from the parametric other ones and may also be applied to nonlinear systems as well ([6],[17]). \\

    When $R_q$ is involutive, the linear differential operator ${\cal{D}}:E\stackrel{j_q}{\rightarrow} J_q(E)\stackrel{\Phi}{\rightarrow} J_q(E)/R_q=F_0$ of order $q$ with space of solutions $\Theta\subset E$ is said to be {\it involutive} and one has the canonical {\it linear Janet sequence} ([17, p 144]):\\
\[  0 \longrightarrow  \Theta \longrightarrow T \stackrel{\cal{D}}{\longrightarrow} F_0 \stackrel{{\cal{D}}_1}{\longrightarrow}F_1 \stackrel{{\cal{D}}_2}{\longrightarrow} ... \stackrel{{\cal{D}}_n}{\longrightarrow} F_n \longrightarrow 0   \]
with {\it Janet bundles} $ F_r={\wedge}^rT^*\otimes J_q(E)/({\wedge}^rT^*\otimes R_q + \delta ({\wedge}^{r-1}T^*\otimes S_{q+1}T^*\otimes E)) $. 
Each operator ${\cal{D}}_{r+1}:F_r \rightarrow F_{r+1}$ is first order involutive as it is induced by $D:{\wedge}^rT^*\otimes J_{q+1}(E) \rightarrow {\wedge}^{r+1}T^*\otimes J_q(E):\alpha\otimes {\xi}_{q+1}\rightarrow d\alpha\otimes {\xi}_q+(-1)^r\alpha\wedge D{\xi}_{q+1}$ and generates the {\it compatibility conditions} (CC) of the preceding one. As the Janet sequence can be cut at any place, {\it the numbering of the Janet bundles has nothing to do with that of the Poincar\'{e} sequence}, contrary to what many people believe in GR.\\   
  Similarly, we have the involutive {\it first Spencer operator} $D_1:C_0=R_q\stackrel{j_1}{\rightarrow}J_1(R_q)\rightarrow J_1(R_q)/R_{q+1}\simeq T^*\otimes R_q/\delta (g_{q+1})=C_1$ of order one induced by $D:R_{q+1}\rightarrow T^*\otimes R_q$. Introducing the {\it Spencer bundles} $C_r={\wedge}^rT^*\otimes R_q/{\delta}({\wedge}^{r-1}T^*\otimes g_{q+1})$, the first order involutive ($r+1$)-{\it Spencer operator} $D_{r+1}:C_r\rightarrow C_{r+1}$ is induced by $D:{\wedge}^{r+1}T^*\otimes R_{q+1} \rightarrow {\wedge}^{r+1}T^*\otimes R_q$ and we obtain the canonical {\it linear Spencer sequence} ([17, p 150]):\\
\[    0 \longrightarrow \Theta \stackrel{j_q}{\longrightarrow} C_0 \stackrel{D_1}{\longrightarrow} C_1 \stackrel{D_2}{\longrightarrow} C_2 \stackrel{D_3}{\longrightarrow} ... \stackrel{D_n}{\longrightarrow} C_n\longrightarrow 0  \]
\noindent
as the Janet sequence for the first order involutive system $R_{q+1}\subset J_1(R_q)$. Introducing the other Spencer bundles $C_r(E)={\wedge}^rT^*\otimes J_q(E)/\delta({\wedge}^{r-1}T^*\otimes S_{q+1}T^*\otimes  E)$ with $C_r \subset C_r(E)$, the linear Spencer sequence is induced by the {\it linear hybrid sequence}: \\
\[  0 \longrightarrow E \stackrel{j_q}{\longrightarrow} C_0(E) \stackrel{D_1}{\longrightarrow} C_1(E) \stackrel{D_2}{\longrightarrow} C_2 \stackrel{D_3}{\longrightarrow} ... \stackrel{D_n}{\longrightarrow } C_n \longrightarrow 0 \]
\noindent
which is at the same time the Janet sequence for $j_q$ and the Spencer sequence for $J_{q+1}(E)\subset J_1(J_q(E))$ ([17, p 153]). Such a sequence projects onto the Janet sequence and we have the following commutative diagram with exact columns: \\
 
 \[ \begin{array}{rcccccccccccl}
 &&&&& 0 &&0&&0&  &0&  \\
 &&&&& \downarrow && \downarrow && \downarrow &    & \downarrow &  \\
  & 0& \rightarrow& \Theta &\stackrel{j_q}{\longrightarrow}&C_0 &\stackrel{D_1}{\longrightarrow}& \circled{$C_1$} &\stackrel{D_2}{\longrightarrow} & C_2 &\stackrel{D_3}{\longrightarrow} ... \stackrel{D_n}{\rightarrow}& C_n &\rightarrow 0 \\
  &&&&& \downarrow & & \downarrow & & \downarrow & &\downarrow &     \\
   & 0 & \rightarrow & E & \stackrel{j_q}{\longrightarrow} & C_0(E) & \stackrel{D_1}{\longrightarrow} & C_1(E) &\stackrel{D_2}{\longrightarrow} & C_2(E) &\stackrel{D_3}{\longrightarrow} ... \stackrel{D_n}{\longrightarrow} & C_n(E) &   \rightarrow 0 \\
   & & & \parallel && \hspace{5mm}\downarrow {\Phi}_0 & &\hspace{5mm} \downarrow {\Phi}_1 & & \hspace{5mm}\downarrow {\Phi}_2 &  & \hspace{5mm}\downarrow {\Phi}_n & \\
   0 \rightarrow & \Theta &\rightarrow & E & \stackrel{\cal{D}}{\longrightarrow} & \circled{$F_0$} & \stackrel{{\cal{D}}_1}{\longrightarrow} & F_1 & \stackrel{{\cal{D}}_2}{\longrightarrow} & F_2 & \stackrel{{\cal{D}}_3}{\longrightarrow} ... \stackrel{{\cal{D}}_n}{\longrightarrow} & F_n & \rightarrow  0 \\
   &&&&& \downarrow & & \downarrow & & \downarrow &   &\downarrow &   \\
   &&&&& 0 && 0 && 0 &&0 &  
   \end{array}     \]
\vspace{5mm}  \\
\noindent   

In this diagram, only depending on the linear differential operator ${\cal{D}}=\Phi\circ j_q$, the epimorhisms ${\Phi}_r:C_r(E)\rightarrow F_r$ for $0\leq r \leq n$ are induced by the canonical projection $\Phi={\Phi}_0:C_0(E)=J_q(E)\rightarrow J_q(E)/R_q=F_0$ if we start with the knowledge of $R_q\subset J_q(E)$ or from the knowledge of an epimorphism $\Phi:J_q(E)\rightarrow F_0$ if we set $R_q=ker(\Phi)$. In the theory of Lie equations coinsidered, $E=T$, $R_q\subset J_q(T)$ is a transitive  involutive system of infinitesimal Lie equations of order $q$ and the corresponding operator $\cal{D}$ is a Lie operator. As an exercise, we invite the reader to draw this diagram in the affine and projective 1-dimensional cases. \\

\noindent
 {\bf EXAMPLE 2.3} : If we restrict our study to the group of isometries of the euclidean metric $\omega$ in dimension $n\geq 2$, exhibiting the Janet and the Spencer sequences is not easy at all, even when $n=2$, because the corresponding Killing operator ${\cal{D}}\xi={\cal{L}}(\xi)\omega=\Omega\in S_2T^*$, involving the Lie derivative ${\cal{L}}$ and providing twice the so-called infinitesimal deformation tensor $\epsilon$ of continuum mechanics, is not involutive. In order to overcome this problem, one must differentiate once by considering also the Christoffel symbols $\gamma$ and add the operator ${\cal{L}}(\xi)\gamma=\Gamma \in S_2T^*\otimes T $. Now, {\it one can prove that the Spencer sequence for Lie groups of transformations is locally isomorphic to the tensor product of the Poincar\'{e} sequence by the Lie algebra of the underlying Lie group}. Hence, if two Lie groups $G'\subset G$ act on $X$, it follows from the definition of the Janet and Spencer bundles that {\it the Spencer sequence for} $G'$ {\it is embedded into the Spencer sequence for} $G$ while {\it the Janet sequence for} $G'$ {\it projects onto the Janet sequence for} $G$ but {\it the common differences are isomorphic to} ${\wedge}^rT^*\otimes ({\cal{G}}/{\cal{G}}')$. This rather philosophical comment, namely to replace the Janet sequence by the Spencer sequence, must be considered as the {\it crucial key} for understanding the work 
of the brothers E. and F. Cosserat in 1909 ([3],[19],[21],[22]) or H. Weyl in 1918 ([22],[26]), the best picture being that of Janet and Spencer playing at see-saw. Indeed, when $n=2$, one has 3 parameters (2 translations + 1 rotation) and the following commutative diagram which only depends on the left commutative square:  \\
  \[  \begin{array}{rccccccccccccr}
 &&&&& 0 &&0&&0&  & \\
 &&&&& \downarrow && \downarrow && \downarrow  &\\
  & 0& \longrightarrow& \Theta &\stackrel{j_2}{\longrightarrow}&\circled{\:3\:}&\stackrel{D_1}{\longrightarrow}&\circled{\:6\:} &\stackrel{D_2}{\longrightarrow} & \circled{\:3\:}&\longrightarrow  0 & \hspace{3mm}Spencer  \\
  &&&&& \downarrow & & \downarrow & & \downarrow & &    & \\
   & 0 & \longrightarrow & \circled{\:2\:} & \stackrel{j_2}{\longrightarrow} & \circled{12}& \stackrel{D_1}{\longrightarrow} & \circled{16} &\stackrel{D_2}{\longrightarrow} & \circled{\:6\:} &   \longrightarrow 0 &\\
   & & & \parallel && \hspace{5mm}\downarrow {\Phi}_0 & &\hspace{5mm} \downarrow {\Phi}_1 & & \hspace{5mm}\downarrow {\Phi}_2 &  &\\
   0 \longrightarrow & \Theta &\longrightarrow & \circled{\:2\:} & \stackrel{\cal{D}}{\longrightarrow} & \circled{\:9\:} & \stackrel{{\cal{D}}_1}{\longrightarrow} & \circled{10} & \stackrel{{\cal{D}}_2}{\longrightarrow} & \circled{\:3\:}& \longrightarrow  0 & \hspace{7mm} Janet \\
   &&&&& \downarrow & & \downarrow & & \downarrow &      &\\
   &&&&& 0 && 0 && 0  &  &
   \end{array}     \]
   In this diagram, there is no way to compare ${\cal{D}}_1$ (curvature alone as in Vessiot) with $D_2$ (curvature + torsion as in Cartan).\\
   
For proving that the adjoint of $D_1$ provides the Cosserat equations which can be parametrized by the adjoint of $D_2$, we may lower the upper indices by means of the constant euclidean metric and look for the factors of ${\xi}_1,{\xi}_2$ and ${\xi}_{1,2}=-{\xi}_{2,1}$ in the integration by parts of the sum:\\
 \[ {\sigma}^{11}({\partial}_1{\xi}_1-{\xi}_{1,1})+{\sigma}^{12}({\partial}_2{\xi}_1-{\xi}_{1,2})+{\sigma}^{21}({\partial}_1{\xi}_2-{\xi}_{2,1})+{\sigma}^{22}({\partial}_2{\xi}_2-{\xi}_{2,2})+{\mu}^{r}({\partial}_r{\xi}_{1,2}-{\xi}_{1,2r}) \]
 in order to obtain: \\
 \[ {\partial}_1{\sigma}^{11}+{\partial}_2{\sigma}^{12}=f^1, \hspace{5mm}   {\partial}_1{\sigma}^{21}+{\partial}_2{\sigma}^{22}=f^2, \hspace{5mm}  {\partial}_1{\mu}^1+{\partial}_2{\mu}^2+{\sigma}^{12}-{\sigma}^{21}=m  \]
Finally, we get the nontrivial {\it first order} parametrization ${\sigma}^{11}={\partial}_2{\phi}^1, {\sigma}^{12}=-{\partial}_1{\phi}^1, {\sigma}^{21}=-{\partial}_2{\phi}^2, {\sigma}^{22}={\partial}_1{\phi}^2, {\mu}^{1}={\partial}_2{\phi}^3+{\phi}^1, {\mu}^{2}=-{\partial}_1{\phi}^3-{\phi}^2$ by means of the three arbitrary functions ${\phi}^1,{\phi}^2,{\phi}^3$, in a coherent way with the Airy {\it second order} parametrization obtained if we set  ${\phi}^1={\partial}_2{\phi}, {\phi}^2={\partial}_1{\phi}, {\phi}^3=-\phi$ when ${\mu}^1=0,{\mu}^2=0$ as we shall see in the third part.\\
    
 The link between the FI of $R_q$ and the CC of $\cal{D}$ is expressed by the following diagram that may be used inductively:  \\   
 \[  \begin{array}{rcccccccl}
      & 0 & & 0 & & 0 & &   \circled{CC}  &  \\
      & \downarrow & & \downarrow & & \downarrow & & &  \\
      0\rightarrow & g_{q+r} & \rightarrow & S_{q+r}T^*\otimes E & \stackrel{{\sigma}_r(\Phi)}{\longrightarrow} & S_rT^*\otimes F_0 & \rightarrow   & coker({\sigma}_r(\Phi)) & \rightarrow 0  \\
        &  \downarrow & & \downarrow & & \downarrow & & \downarrow   &  \\
 0 \rightarrow & R_{q+r} & \rightarrow & J_{q+r}(E) &\stackrel{{\rho}_r(\Phi)}{ \longrightarrow }  & J_r(F_0) & \rightarrow & coker({\rho}_r(\Phi)) & \rightarrow 0  \\
     &  \downarrow & &  \hspace{11mm} \downarrow {\pi}^{q+r}_{q+r-1} & & \hspace{8mm} \downarrow {\pi}^r_{r-1} & & \downarrow &   \\ 
 0 \rightarrow & R_{q+r-1} & \rightarrow & J_{q+r-1}(E) & \stackrel{ {\rho}_{r-1}(\Phi)}{ \longrightarrow } & J_{r-1}(F_0) & \rightarrow & coker({\rho}_{r-1}(\Phi) & \rightarrow 0  \\ 
    &  & & \downarrow & & \downarrow & & \downarrow    & \\
    & \circled{\,FI\,} &  &  0  &  &  0  &  & 0  &    
    \end{array}   \]
The " {\it snake theorem} " ([17],[23]) then provides the long exact {\it connecting sequence}:  \\   
\[ 0 \rightarrow g_{q+r} \rightarrow  R_{q+r} \rightarrow R_{q+r-1} \rightarrow  coker({\sigma}_r(\Phi)) \rightarrow coker({\rho}_r(\Phi)) \rightarrow coker({\rho}_{r-1}(\Phi))  \rightarrow 0  \]

If we apply such a diagram to first order Lie equations with no zero or first order CC, we have $q=1, E=T$ and we may apply the Spencer $\delta$-map to the top row with $r=2$ in order to get the commutative diagram:  \\    
\[  \begin{array}{rcccccccl}
   &  0 & & 0 & & 0 &  &  &   \\
   & \downarrow & & \downarrow & & \downarrow & & &  \\
0\rightarrow & g_3 & \rightarrow &  S_3T^*\otimes T & \rightarrow & S_2T^*\otimes F_0& \rightarrow & F_1 & \rightarrow 0  \\
   & \hspace{2mm}\downarrow  \delta  & & \hspace{2mm}\downarrow \delta & &\hspace{2mm} \downarrow \delta & & &  \\
0\rightarrow& T^*\otimes g_2&\rightarrow &T^*\otimes S_2T^*\otimes T & \rightarrow &T^*\otimes T^*\otimes F_0 &\rightarrow & 0 &  \\
   &\hspace{2mm} \downarrow \delta &  &\hspace{2mm} \downarrow \delta & &\hspace{2mm}\downarrow \delta &  &  &   \\
0\rightarrow & {\wedge}^2T^*\otimes g_1 & \rightarrow & \underline{{\wedge}^2T^*\otimes T^*\otimes T} & \rightarrow & {\wedge}^2T^*\otimes F_0 & \rightarrow & 0 &  \\
   &\hspace{2mm}\downarrow \delta  &  & \hspace{2mm} \downarrow \delta  &  & \downarrow  & &  &  \\
0\rightarrow & {\wedge}^3T^*\otimes T & =  & {\wedge}^3T^*\otimes T  &\rightarrow   & 0  &  &  &   \\
    &  \downarrow  &  &  \downarrow  &  &  &  &  &  \\
    &  0  &   & 0  & &  &  &  &
\end{array}  \]
with exact rows and exact columns but the first that may not be exact at ${\wedge}^2T^*\otimes g_1$. We shall denote by $B^2(g_1)$ the {\it coboundary} as the image of the central $\delta$, by $Z^2(g_1)$ the {\it cocycle} as the kernel of the lower $\delta$ and by $H^2(g_1)$ the {\it Spencer} $\delta$-{\it  cohomology} at ${\wedge}^2T^*\otimes g_1$ as the quotient.  \\

In the classical Killing system, $g_1\subset T^*\otimes T$ is defined by ${\omega}_{rj}(x){\xi}^r_i+{\omega}_{ir}(x){\xi}^r_j=0 \Rightarrow {\xi}^r_r=0, g_2=0,g_3=0$. Applying the previous diagram, we discover that the Riemann tensor is a section of the bundle $ Riemann=F_1=H^2(g_1)=Z^2(g_1)$ with $dim(Riemann)= (n^2(n+1)^2/4)-(n^2(n+1)(n+2)/6)=(n^2(n-1)^2/4)-(n^2(n-1)(n-2)/6)=n^2(n^2-1)/12$ by using the top row or the left column. Though we discover the two properties of the Riemann tensor through the chase involved, we have no indices and cannot therefore exhibit the {\it Ricci tensor} of GR by means of the usual {\it contraction} or {\it trace}.  \\
 
 Let us proceed the same way with the {\it conformal Killing system} ${\hat{\Omega}}_{ij}\equiv ({\cal{L}}(\xi){\hat{\omega}})_{ij}\equiv {\hat{\omega}}_{rj}{\partial}_i{\xi}^r+{\hat{\omega}}_{ir}{\partial}_j{\xi}^r-\frac{2}{n}{\omega}_{ij}{\partial}_r{\xi}^r+{\xi}^r{\partial}_r{\hat{\omega}}_{ij}=0 $ obtained by introducing ${\hat{\omega}}_{ij}={\omega}_{ij}/ {\mid det(\omega)\mid}^{\frac{1}{n}}$ or, equivalently, by eliminating $A(x)$ in ${\cal{L}}(\xi)\omega=A(x)\omega$. Now ${\hat{g}}_1$ is defined by ${\omega}_{rj}{\xi}^r_i+{\omega}_{ir}{\xi}^r_j-\frac{2}{n}{\omega}_{ij}{\xi}^r_r=0$ but we have ${\hat{g}}_3=0, \forall n\geq 3$ with $H^2({\hat{g}}_2)=0, \forall n\geq 4$ and the {\it Weyl tensor} is a section of the bundle $Weyl={\hat{F}}_1=H^2({\hat{g}}_1)= Z^2({\hat{g}}_1)/\delta(T^*\otimes {\hat{g}}_2)$ with $dim(Weyl)=n(n+1)(n+2)(n-3)/12$. Similarly, we have no indices and cannot therefore exhibit the Ricci tensor. However, when $n=4$, among the components of the Spencer operator we have ${\partial}_i{\xi}^r_{rj}- {\xi}^r_{rij}={\partial}_i{\xi}^r_{rj}$ and thus ${\partial}_i{\xi}^r_{rj}-{\partial}_j{\xi}^r_{ri}=F_{ij}$. Such a result allows to recover the electromagnetic (EM) field in the image of the Spencer operator $D_1$ and Maxwell equations by duality along the way proposed by Weyl in ([26]) but {\it the use of the Spencer operator provides the only possibility to exhibit a link with Cosserat 
 equations}.   \\ 
 
 Comparing the classical and conformal Killing systems by using the inclusions $R_1\subset {\hat{R}}_1 \Rightarrow g_1\subset {\hat{g}}_1$, we finally obtain the following commutative and exact diagram where a diagonal chase allows to identify $Ricci$ with $S_2T^*\subset T^*\otimes T^*\simeq T^*\otimes {\hat{g}}_2 $ and {\it it split} the right column ([22]):   \\
 \[ \begin{array}{rcccccccll}
 & & & & & & & 0 & &\\
 & & & & & & & \downarrow && \\
  & & & & & 0& & Ricci & & \\
  & & & & & \downarrow & & \downarrow  & & \\
   & & & 0 &\rightarrow & Z^2(g_1) & \rightarrow & Riemann  & \rightarrow 0 & \\
   & & & \downarrow & & \downarrow & &  \downarrow  &  &JANET\\
   & 0 &\rightarrow & T^*\otimes {\hat{g}}_2 & \stackrel{\delta}{\rightarrow} & Z^2({\hat{g}}_1) & \rightarrow & Weyl & \rightarrow 0 & \\
    & & & \downarrow & & \downarrow & & \downarrow &  & \\
 0 \rightarrow & S_2T^* & \stackrel{\delta}{\rightarrow}& T^*\otimes T^* &\stackrel{\delta}{\rightarrow} & {\wedge}^2T^* & \rightarrow & 0 &  & \\
   & & & \downarrow &  & \downarrow & & & & \\
   & & & 0 & & 0 & & &  &\\
   &&&& SPENCER &&&&&
   \end{array}  \]
\vspace*{5mm}  \\  

\noindent
{\bf SECOND FUNDAMENTAL RESULT}: The Ricci tensor only depends on the "difference" existing between the clasical Killing system and the conformal Killing system, namely the $n$ second order jets ({\it elations} once more). The Ricci tensor, thus obtained without contracting the indices as usual, may be embedded in the image of the Spencer operator made by $1$-forms with value in $1$-forms that we have already exhibited for describing EM. It follows that the foundations of both EM and GR are not coherent with jet theory and must therefore be revisited within this new framework.  \\

\noindent
{\bf THIRD PART: ALGEBRAIC ANALYSIS}:  \\

\noindent
{\bf EXAMPLE 3.1}: Let a rigid bar be able to slide along an horizontal axis with reference position $x$ and attach two pendula, one at each end, with lengths $l_1$ and $l_2$, having small angles ${\theta}_1$ and ${\theta}_2$ with respect to the vertical. If we project Newton law with gravity $g$ on the perpendicular to each pendulum in order to eliminate the tension of the threads and denote the time derivative with a dot, we get the two equations:  \\
\[ \ddot{x}+l_1{\ddot{\theta}}_1+g{\theta}_1=0 , \hspace{2cm}  \ddot{x}+l_2{\ddot{\theta}}_2+g{\theta}_2=0  \]
As an {\it experimental fact}, starting from an arbitrary movement of the pendula, we can stop them if and only if $l_1\neq l_2$ and we say that the system is {\it controllable}. \\

More generally, we can bring the OD equations describing the behaviour of a mechanical or electrical system to the {\it Kalman form} $\dot{y}=Ay+bu$ with {\it input} $u=(u^1, ... , u^p)$ and {\it output} $y=(y^1, ..., y^m)$. We say that the system is controllable if, for any given $y(0),y(T),T<\infty$, one can find $u(t)$ such that a coherent {\it trajectory} $y(t)$ may be found. In 1963, R.E. Kalman discovered that the system is controllable if and only if $rk(B,AB,...,A^{m-1}B)=m$. Surprisingly, such a {\it functional definition} admits a {\it formal test} which is only valid for Kalman type systems with constant coefficients and is thus far from being intrinsic. In the PD case, the Spencer form will replace the Kalman form.  \\

\noindent
{\bf EXAMPLE 3.2}: $\dot{y}-\dot{u}=0 \Rightarrow  y(t)-u(t)=c=cst \Rightarrow u(T)-u(0)=y(T)-y(0)$ can always be achieved and the system is thus controllable in the sense of the definition but $z=y-u \Rightarrow \dot{z}=0$ is not controllable in the sense of the test.  \\

\noindent
{\bf EXAMPLE 3.3}: \hspace{1cm}  ${\dot{y}}^1-a(t)y^2-{\dot{y}}^3=0, \hspace{5mm} y^1-{\dot{y}}^2+{\dot{y}}^3=0  $.  \\
{\it Any way} to bring this system to Kalman form provides the controllability condition $a(a-1)\neq 0$ if $a=cst$ but nothing can be said if $a=a(t)$. Also, getting $y^1$ from the second equation and substituting in the first, we get the second order OD equation ${\ddot{y}}^2-{\ddot{y}}^3-{\dot{y}}^3-a(t)y^2=0$ for which nothing can be said at first sight.  \\

\noindent
{\bf PROBLEM 1}: Is a SYSTEM of OD or PD equations " {\it controllable} " (answer must be YES or NO) and how can we define controllability ?.  \\

 Now, if a differential operator $\xi \stackrel{\cal{D}}{\longrightarrow} \eta$ is given, a {\it direct problem} is to find (generating) {\it compatibility conditions} (CC) as an operator $\eta \stackrel{{\cal{D}}_1}{\longrightarrow} \zeta $ such that ${\cal{D}}\xi=\eta \Rightarrow {\cal{D}}_1\eta=0$. Conversely, the {\it inverse problem} will be to find $\theta \stackrel{{\cal{D}}_{-1}}{\longrightarrow} \xi$ such that $\cal{D}$ generates the CC of ${\cal{D}}_{-1}$ and we shall say that $\cal{D}$ is {\it parametrized} by ${\cal{D}}_{-1}$. Of course, solving the direct problem (Janet, Spencer) is {\it necessary} for solving the inverse problem.  \\
 
 \noindent
 {\bf EXAMPLE 3.4}: When $n=2$, the Cauchy equations for the stress in continuum mechanics are ${\partial}_1{\sigma}^{11}+{\partial}_2{\sigma}^{21}=0, {\partial}_1{\sigma}^{12}+{\partial}_2{\sigma}^{22}=0$ with ${\sigma}^{12}={\sigma}^{21}$. Their parametrization ${\sigma}^{11}={\partial}_{22}\phi, {\sigma}^{12}={\sigma}^{21}=-{\partial}_{12}\phi, {\sigma}^{22}={\partial}_{11}\phi$  has been discovered by Airy in 1862 and $\phi$ is called the {\it Airy function}. When $n=3$, Maxwell and Morera discovered a similar parametrization with 3 potentials (exercise).  \\
 
 \noindent
 {\bf EXAMPLE 3.5}: When $n=4$, the Maxwell equations $dF=0$ where $F\in {\wedge}^2T^*$ is the EM field are parametrized by $dA=F$ where $A\in T^*$ is the $4$-potential. The second set of Maxwell equations can also be parametrized by the so-called pseudopotential which is a pseudovector density (exercise).  \\

\noindent
{\bf EXAMPLE 3.5}: If $n=4$, $\omega$ is the Minkowski metric and $\phi=GM/r$ is the gravitational potential, then $\phi/c^2\ll 1$ and a perturbation $\Omega$  of $\omega$ may satisfy in vacuum the $10$ second order {\it Einstein equations} for the $10$ $ \Omega$:  \\
\[ {\omega}^{rs}(d_{ij}{\Omega}_{rs}+d_{rs}{\Omega}_{ij}-d_{ri}{\Omega}_{sj}-d_{sj}{\Omega}_{ri})-{\omega}_{ij}({\omega}^{rs}{\omega}^{uv}d_{rs}{\Omega}_{uv}-{\omega}^{ru}{\omega}^{sv}d_{rs}{\Omega}_{uv})=0  \]
The parametrizing challenge has been proposed in 1970 by J. Wheeler for 1000 \$ and solved {\it negatively} in 1995 by the author who only received 1 \$. \\

\noindent
{\bf PROBLEM 2}: Is an OPERATOR parametrizable (answer must be YES or NO) and how can we find a parametrization ?.  \\

Let $A$ be a {\it unitary ring}, that is $1,a,b\in A \Rightarrow a+b,ab \in A, 1a=a$ and even an {\it integral domain}, that is $ab=0\Rightarrow a=0$ or $b=0$. We say that $M={}_AM$ is a {\it left module} over $A$ if $x,y\in M\Rightarrow ax,x+y\in M, \forall a\in A$ and we denote by $hom_A(M,N)$ the set of morphisms $f:M\rightarrow N$ such that $f(ax)=af(x)$.  \\

\noindent
{\bf DEFINITION 3.6}: We define the {\it torsion submodule} $t(M)=\{x\in M\mid \exists 0\neq a\in A, ax=0\}\subseteq M$.  \\

There is a sequence $0 \rightarrow t(M) \rightarrow M \stackrel{\epsilon}{\longrightarrow} hom_A(hom_A(M,A),A)  $ where the morphism $\epsilon$ is defined by $\epsilon(x)(f)=f(x), \forall x\in M, \forall f\in hom_A(M,A)$ because we have at once $af(x)=f(ax)=f(0)=0 \Rightarrow t(M)\subseteq ker(\epsilon)$. \\

\noindent
{\bf PROBLEM 3}: Is a MODULE $M$ torsion-free, that is $t(M)=0$ (answer must be YES or NO) and how can we test such a property ?.  \\

In the remaining of this paper we shall prove that the three problems are indeed identical and that {\it only the solution of the third will provide the solution of the two others}. \\

Let $\mathbb{Q}\subset K$ be a {\it differential field}, that is a field ($a\in K \Rightarrow 1/a\in K$) with $n$ commuting {\it derivations} $\{{\partial}_1,...,{\partial}_n\}$ with ${\partial}_i{\partial}_j={\partial}_j{\partial}_i={\partial}_{ij}, \forall i,j=1,...,n$ such that ${\partial}_i(a+b)={\partial}_ia+{\partial}_ib$ and ${\partial}_i(ab)=({\partial}_ia)b+a{\partial}_ib, \forall a,b\in K$. Using an implicit summation on multiindices, we may introduce the (noncommutative) {\it ring of differential operators} $D=K[d_1,...,d_n]=K[d]$ with elements $P=a^{\mu}d_{\mu}$ such that $\mu<\infty$ and $d_ia=ad_i+{\partial}_ia$. Now, if we introduce {\it differential indeterminates} $y=(y^1,...,y^m)$, we may extend $d_iy^k_{\mu}=y^k_{\mu+1_i}$ to ${\Phi}^{\tau}\equiv a^{\tau\mu}_ky^k_{\mu}\stackrel{d_i}{\longrightarrow} d_i{\Phi}^{\tau}\equiv a^{\tau\mu}_ky^k_{\mu+1_i}+{\partial}_ia^{\tau\mu}_ky^k_{\mu}$ for $\tau=1,...,p$. Therefore, setting $Dy^1+...+dy^m=Dy\simeq D^m$, we obtain by residue the {\it differential module} or $D$-{\it module} $M=Dy/D\Phi$. Introducing the two free differential modules $F_0\simeq D^m, F_1\simeq D^p$, we obtain equivalently the {\it free presentation} $F_1\stackrel{\cal{D}}{\rightarrow} F_0 \rightarrow M \rightarrow 0$. More generally, introducing the successive CC as in the preceding section, we may finally obtain the {\it free resolution} of $M$, namely the exact sequence $\hspace{5mm} ... \stackrel{{\cal{D}}_2}{\longrightarrow} F_2  \stackrel{{\cal{D}}_1}{\longrightarrow} F_1 \stackrel{\cal{D}}{\longrightarrow}F_0\longrightarrow M \longrightarrow 0 $.  \\                                  
  The " {\it trick} " is to let $\cal{D}$ act on the left on column vectors in the operator case and on the right on row vectors in the module case. Homological algebra has been created for finding intrinsic properties of modules not depending on their presentation or even on their resolution.  \\

\noindent
{\bf EXAMPLE 3.7}: In order to understand that different presentations may nevertheless provide isomorphic modules, let us consider the linear inhomogeneous system $Py\equiv d_{22}y=u, Q\equiv d_{12}y-y=v $ with $K=\mathbb{Q}$. Differentiating twice, we get $y=d_{11}u-d_{12}v-v$ and the two fourth order CC:  \\
\[   A\equiv d_{1122}u-d_{1222}v-d_{22}v-u=0, \hspace{1cm} B\equiv d_{1112}u-d_{1122}v-d_{11}u=0  \]
However, as $PQ=QP$, we also get the CC $ C\equiv d_{12}u-d_{22}v-u=0$ and the two resolutions:  \\
\[   0 \longrightarrow D \longrightarrow D^2 \longrightarrow D^2 \longrightarrow M \longrightarrow 0, \hspace{1cm} 0 \longrightarrow D \longrightarrow D^2 \longrightarrow M \longrightarrow 0 \]
where we have identified the two differential modules involved on the right because:  \\
\[   A\equiv d_{12}C+C, \hspace{5mm} B\equiv d_{11}C \hspace{5mm}\Leftrightarrow \hspace{5mm} C\equiv d_{22}B-d_{12}A+A  \]
\hspace*{3mm}  We now exhibit another approach by defining the {\it formal adjoint} of an operartor $P$ and an operator matrix $\cal{D}$:  \\

\noindent
{\bf DEFINITION 3.8}: \hspace{1cm}$P=a^{\mu}d_{\mu}\in D  \stackrel{ad}{\longleftrightarrow} ad(P)=(-1)^{\mid\mu\mid}d_{\mu}a^{\mu}   \in D $  \\
\[ <\lambda,{\cal{D}} \xi>=<ad({\cal{D}})\lambda,\xi>+\hspace{1mm} {div}\hspace{1mm} ( ... )  \]
from integration by part, where $\lambda$ is a row vector of test functions and $<  > $ the usual contraction.  \\

\noindent
{\bf PROPOSITION 3.9}: If we have an operator $E\stackrel{\cal{D}}{\longrightarrow} F$, we obtain by duality an operator ${\wedge}^nT^*\otimes E^*\stackrel{ad(\cal{D})}{\longleftarrow} {\wedge}^nT^*\otimes F^*$where $E^*$ is obtained from $E$ by inverting the transition matrix and EM provides a fine example of such a procedure ([12]).  \\

Now, with operational notations, let us consider the two differential sequences:  \\
\[   \xi  \stackrel{\cal{D}}{\longrightarrow} \eta \stackrel{{\cal{D}}_1}{\longrightarrow} \zeta  \]
\[   \nu  \stackrel{ad(\cal{D})}{\longleftarrow} \circled{$\mu$} \stackrel{ad({\cal{D}}_1)}{\longleftarrow} \lambda   \]
where ${\cal{D}}_1$ generates all the CC of $\cal{D}$. Then ${\cal{D}}_1\circ {\cal{D}}\equiv 0 \Longleftrightarrow ad({\cal{D}})\circ ad({\cal{D}}_1)\equiv 0 $ but $ad(\cal{D})$ may not generate all the CC of $ad({\cal{D}}_1)$.   \\

\noindent
{\bf EXAMPLE 3.10}: With ${\partial}_{22}\xi={\eta}^2, {\partial}_{12}\xi={\eta}^1$ for $\cal{D}$, we get  ${\partial}_1{\eta}^2-{\partial}_2{\eta}^1=\zeta$ for ${\cal{D}}_1$. Then $ad({\cal{D}}_1)$ is defined by ${\mu}^2=-{\partial}_1\lambda, {\mu}^1={\partial}_2\lambda$ while $ad(\cal{D})$ is defined by $\nu={\partial}_{12}{\mu}^1+{\partial}_{22}{\mu}^2$ but the CC of $ad({\cal{D}}_1)$ are generated by ${\nu}'={\partial}_1{\mu}^1+{\partial}_2{\mu}^2$. Passing to the module framework, we obtain the sequences:  \\
\[  \begin{array}{rccccl}
 D & \stackrel{{\cal{D}}_1}{\longrightarrow} & D^2 & \stackrel{\cal{D}}{\longrightarrow} & D & \longrightarrow M \longrightarrow  0  \\
  D& \stackrel{ad({\cal{D}}_1)}{\longleftarrow} & \circled{$D^2$} & \stackrel{ad(\cal{D})}{\longleftarrow} & D &  
  \end{array}  \]

\noindent
{\bf THEOREM 3.11}: The cohomology $ext^1(M)$ at $D^2$ of the lower sequence does not depend on the resolution of $M$ and is a torsion module called the {\it first extension module} of $M$.  \\

  Exactly like we defined the differential module $M$ from $\cal{D}$, let us define the differential module $N$ from $ad(\cal{D})$. The proof of the next theorem is quite tricky and out of the scope of this paper ([8],[18],[20]): \\

\noindent
{\bf MAIN THEOREM 3.12}: \hspace{2cm}  $  ext^1(N)=t(M)=ker(\epsilon)  $   \\

\noindent
{\bf FORMAL TEST 3.13}: The {\it double duality test} needed in order to check whether $t(M)=0$ or not and to find out a parametrization if $t(M)=0$ has 5 steps which are drawn in the following diagram where $ad({\cal{D}}_{-1})$ generates the CC of $ad(\cal{D})$ and ${\cal{D}}'$ generates the CC of ${\cal{D}}_{-1}$:  \\
\[  \begin{array}{rcccccl}
 & & & & &  {\eta}' &\hspace{1cm} \circled{5}  \\
 & & & & \stackrel{{\cal{D}}'}{\nearrow} &  &  \\
\circled{4} \hspace{1cm}& \theta  & \stackrel{{\cal{D}}_{-1}}{\longrightarrow} &  \xi & \stackrel{\cal{D}}{\longrightarrow} & \eta &\hspace{1cm}   \circled{1}  \\
 &  &  &  &  &  &  \\
 &  &  &  &  &  &  \\
 \circled{3} \hspace{1cm}& \nu & \stackrel{ad({\cal{D}}_{-1})}{\longleftarrow} & \mu & \stackrel{ad(\cal{D})}{\longleftarrow} & \lambda &\hspace{1cm} \circled{2}
  \end{array}  \]
\vspace{2mm}  \\

\noindent
{\bf THEOREM 3.14}: $\cal{D}$ parametrized by ${\cal{D}}_{-1} \Leftrightarrow {\cal{D}}={\cal{D}}' \Leftrightarrow t(M)=0 \Leftrightarrow ext^1(N)=0 $.  \\

\noindent
{\bf COROLLARY 3.15}: If $n=1$ and $\cal{D}$ is {\it surjective}, then $t(M)=0$ if and only if $ad(\cal{D})$ is injective ([10,p211],[20]).  \\

\noindent
{\bf EXAMPLE 3.16}: ({\it Kalman test revisited}) If we multiply the Kalman system $-\dot{y}+Ay+Bu=0$ on the left by a test row vector $\lambda$, we obtain:  \\
\[  ad(\cal{D})\hspace{5mm} \left \{
\begin{array}{rcr}
y & \longrightarrow  &  \dot{\lambda}+\lambda A=0  \\
u &  \longrightarrow &   \lambda B=0
\end{array}
\right.      \]
Differentiating the zero order equations and using the first order ones, we get $\lambda AB=0$ and so on. Using the Cayley-Hamilton theorem, we stop at $\lambda A^{m-1}B=0$ and find back exactly the Kalman test but in a completely different intrinsic framework.  \\

\noindent
{\bf EXAMPLE 3.17}: ({\it Double pendulum revisited}) Using two test functions ${\lambda}_1$ and ${\lambda}_2$, we get:\\
\[  ad(\cal{D})  \hspace{5mm} \left\{
\begin{array}{rcr}
x \hspace{1mm}& \longrightarrow & {\ddot{\lambda}}^1+{\ddot{\lambda}}^2 =0  \\
{\theta}_1 & \longrightarrow & l_1{\ddot{\lambda}}^1+g{\lambda}^1=0  \\
{\theta}_2  & \longrightarrow & l_2{\ddot{\lambda}}^2+g{\lambda}^2=0
\end{array}   
\right.   \]
and obtain at once the zero order equation $l_1{\lambda}^2+l_2{\lambda}^1=0$. Differentiating twice and substituting, we also get $(l_1/l_2){\lambda}^2+(l_2/l_1){\lambda}^1=0$ and $ad(\cal{D})$ is injective if and only if $l_2-l_1\neq 0$.  \\

\noindent
{\bf EXAMPLE 3.18}:({\it Airy parametrization revisited}) When $n=2$, we may study the {\it infinitesimal deformation} $\epsilon\in S_2T^*$ by means of the {\it Killing operator} ${\cal{D}}\xi\equiv {\cal{L}}(\xi)\omega=2\epsilon$ when $\omega$ is the euclidean metric. Then $ad(\cal{D})$ provides (up to sign and factor $2$) the {\it Cauchy equations}  ${\partial}_r{\sigma}^{rj}+{\gamma}^j_{rs}{\sigma}^{rs}=0$ for the {\it stress tensor density} ([16],[26]). The following diagram describes the {\it Poincar\'e scheme}:  \\
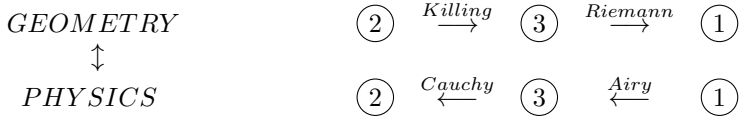
\[  \begin{array}{rccccl}
GEOMETRY \hspace{20mm} &\circled{2} & \stackrel{Killing}{\longrightarrow} &\circled{3} & \stackrel{Riemann}{\longrightarrow} &\circled{1}\hspace{45mm}  \\
\updownarrow \hspace{30mm} &  &  &  &  &  \\
PHYSICS \hspace{23mm} & \circled{2} & \stackrel{Cauchy}{\longleftarrow} & \circled{3} & \stackrel{Airy}{\longleftarrow} & \circled{1} \hspace{45mm} 
\end{array}  \]
Accordingly, the second order {\it Airy parametrization} is nothing else than the adjoint of the only Riemann CC involved, namely ${\partial}_{11}{\epsilon}_{22}+{\partial}_{22}{\epsilon}_{11}-2{\partial}_{12}{\epsilon}_{12}=0$ which is the linearization of the Riemann tensor of Example 1.9 .\\

\noindent
{\bf EXAMPLE 3.19}: ({\it Einstein equations revisited}) Contrary to the Ricci operator, the Einstein operator is self-adjoint because it comes from a variational procedure, the sixth terms being exchanged between themselves under $ad$. For example, we have:  \\
\[  {\lambda}^{ij}({\omega}^{rs}d_{ij}{\Omega}_{rs}) \stackrel{ad}{\leftrightarrow} ({\omega}^{rs}d_{ij}{\lambda}^{ij}){\Omega}_{rs}=({\omega}_{ij}d_{rs}{\lambda}^{rs}){\Omega}^{ij}  \]
and the adjoint of the first operator is the sixth. Accordingly, one has the following diagram where ${\cal{D}}\neq {\cal{D}}'$:  \\
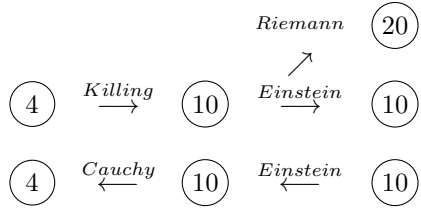
\[  \begin{array}{rcccl}
  &  &  &\stackrel{Riemann}{ }  & \circled{20}   \\
  & &  & \nearrow &    \\
 \circled{\:4\:} &  \stackrel{Killing}{\longrightarrow} & \circled{10} & \stackrel{Einstein}{\longrightarrow} & \circled{10}  \\
  & & & &  \\
 \circled{\:4\:} & \stackrel{Cauchy}{\longleftarrow} & \circled{10} & \stackrel{Einstein}{\longleftarrow} & \circled{10}  
\end{array}  \]
\vspace{3mm}  \\
\noindent
{\bf THIRD FUNDAMENTAL RESULT}: Comparing this diagram to the previous one proves that Einstein equations are not coherent with Janet and Spencer sequences as conformal geometry has not been introduced in this last part.  \\

\noindent
{\bf EXERCISE 3.20}: Prove that ${\ddot{y}}^2-{\ddot{y}}^3-{\dot{y}}^3-a(t)y^2=0$ is controllable if and only if $\dot{a}+a^2-a \neq 0$ (Riccati) and find a parametrization.  \\

\noindent
{\bf EXERCISE 3.21}: Prove that the infinitesimal contact transformations of Example 1.10 admit the injective parametrization $-x^3{\partial}_3\theta + \theta={\xi}^1, -{\partial}_3\theta={\xi}^2, {\partial}_2\theta-x^3{\partial}_1\theta={\xi}^3 \Rightarrow {\xi}^1-x^3{\xi}^2=\theta $.  \\

\noindent
{\bf REFERENCES}:\\

\noindent
[1] E. CARTAN, "Sur la structure des groupes infinis de transformations,"  Ann. Ec. Norm. ,21,1904, pp153-206.\\
\noindent
[2] E. CARTAN, "Sur les vari\'{e}t\'{e}s \`{a} connexion affine et la th\'{e}orie de la relativit\'{e} g\'{e}n\'{e}ralis\'{e}e," Ann. Ec. Norm. Sup., 40, 1923, 325-412; 41, 1924, 1-25; 42, 1925, 17-88.\\
\noindent
[3] E. COSSERAT, F. COSSERAT, "Th\'{e}orie des Corps D\'{e}formables," Hermann, Paris, 1909.\\
\noindent
[4] L.P. EISENHART, "Riemannian Geometry," Princeton University Press, Princeton, 1926.\\
\noindent
[5] H. GOLDSCHMIDT, "Sur la structure des \'{e}quations de Lie," J. Differential Geometry, 6, 1972, 357-373 and 7, 1972, 67-95.\\
\noindent
[6] M. JANET, "Sur les syst\`{e}mes aux d\'{e}riv\'{e}es partielles," Journal de Math., 8, (3), 1920, 65-151. \\
\noindent
[7] R.E. KALMAN, Y.C. YO, K.S. NARENDA, "Controllability of linear dynamical systems," Contrib. Diff. Equations, 1 (2), 1963, 189-213.\\
\noindent
[8] M. KASHIWARA, "Algebraic Study of Systems of Partial Differential Equations," M\'emoires de la Soci\'et\'e Math\'ematique de France 63, 1995, 
(Transl. from Japanese of his 1970 Master's Thesis).\\
\noindent
[9] A. KUMPERA, D.C. SPENCER, "Lie Equations, " Ann. Math. Studies 73," Princeton University Press, Princeton, 1972.\\
\noindent
[10] E. KUNZ, "Introduction to Commutative Algebra and Algebraic Geometry," Birkh\"{a}user, 1985.\\
\noindent
[11] A. LORENZ, "Jet Groupoids, Natural Bundles and the Vessiot Equivalence Method," Ph.D. Thesis, Department of Mathematics,  RWTH Aachen-University, march 18, 2009.See also: On local integrability conditions of jet groupoids, Acta Appl. Math., 01, 2008, 205-213, available at :\\
http://dx.doi.org/10.1007/s10440-008-9193-7   \\
\noindent
[12] V. OUGAROV, "Th\'{e}orie de la Relativit\'{e} Restreinte," MIR, Moscow, 1969; french translation, 1979.\\
\noindent
[13] V.P. PALAMODOV, "Linear Differential Operators with Constant Coefficients," 
Grundlehren der Mathematischen Wissenschaften 168, Springer, 1970.\\
\noindent
[14] W. PAULI, "Theory of Relativity," Pergamon Press, London, 1958.\\
\noindent
[15] J.-F. POMMARET, "Differential Galois Theory,"  Gordon and Breach, New York, 1983. \\
\noindent
[16] J.-F. POMMARET, "Lie Pseudogroups and Mechanics,"  Gordon and Breach, New York, 1988.\\
\noindent
[17] J.-F. POMMARET, "Partial Differential Equations and Group Theory," Kluwer, 1994.\\
\noindent
[18] J.-F. POMMARET, "Partial Differential Control Theory," Kluwer, Dordrecht, 2001.\\
\noindent
[19] J.-F. POMMARET, "Group interpretation of Coupling Phenomena," Acta Mechanica, 149, 2001, 23-39.\\
\noindent
[20] J.-F. POMMARET, "Algebraic Analysis of Control Systems Defined by Partial Differential Equations," in: Advanced Topics in Control Systems Theory, Lecture Notes in Control and Information Sciences 311, Chapter 5, Springer, 2005, 155-223.\\
\noindent
[21] J.-F. POMMARET, "Parametrization of Cosserat equations," Acta Mechanica, 215, 2010, 43-55.\\
\noindent
[22]  J.-F. POMMARET, "Spencer Operator and Applications: From Continuum Mechanics to Mathematical Physics," in: Continuum Mechanics-Progress in Fundamentals and Engineering Applications, Dr. Yong Gan (Ed.), ISBN: 978-953-51-0447--6, InTech, 2012, Available from: \\
http://www.intechopen.com/books/continuum-mechanics-progress-in-fundamentals-and-engineerin-applications/spencer-operator-and-applications-from-continuum-mechanics-to-mathematical-physics  \\
\noindent
[23] J. J. ROTMAN, "An Introduction to Homological Algebra," Academic Press, 1979.\\
\noindent
[24] D. C. SPENCER, "Overdetermined Systems of Partial Differential Equations," Bull. Am. Math. Soc., 75, 1965, 1-114.\\
\noindent
[25] E. VESSIOT, "Sur la th\'{e}orie des groupes infinis," Ann. Ec. Norm. Sup., 20, 1903, 411-451.\\
\noindent
[26]  H. WEYL, "Space, Time, Matter," Berlin, 1918 (1922, 1958; Dover, 1952). \\

\end{document}